\documentclass[%
longbibliography,%
amsmath,amssymb,%
showkeys,
reprint,%
]{revtex4-1}
\usepackage{graphicx}
\usepackage{dcolumn}
\usepackage{bm}

\usepackage[utf8]{inputenc}
\usepackage[T1]{fontenc}
\usepackage{mathptmx}
\usepackage{etoolbox}
\usepackage{color,soul}
\usepackage{xcolor}
\usepackage{lineno}
\usepackage{url}
\usepackage[colorlinks=true,linkcolor=blue, citecolor=blue, urlcolor = blue]{hyperref}

\begin{document}

\title{Near-field diffraction of protons by a nanostructured metallic grating under external electric field: Asymmetry and sidebands in Talbot self-imaging}

\author{Sushanta Barman}
\email{sushanta@iitk.ac.in}
\author{Sudeep Bhattacharjee}%
 \email{sudeepb@iitk.ac.in}
\affiliation{ 
	Department of Physics, Indian Institute of Technology-Kanpur, Kanpur 208016, India}%

\begin{abstract}
	
Self-imaging in near-field diffraction is a practical application of coherent manipulation of matter waves in Talbot interferometry. In this work, near-field diffraction of protons by a nanostructured metallic grating under the influence of (a) uniform, (b) spatially modulated, and (c) temporally modulated electric fields are investigated. Time-domain simulations of two-dimensional Gaussian wave packets for protons are performed by solving the time-dependent Schrödinger's equation using the generalized finite difference time domain (GFDTD-Q) method for quantum systems. Effects of strength ($E_0$) and orientation ($\theta$) of the uniform electric field on the diffraction properties, such as fringe pattern, intensity of the peaks, fringe shift, and visibility, are investigated. The results show that the Talbot fringes shift significantly in the transverse direction even for a small change in the applied electric field ($\Delta E_0$ $=0.1$ V/m) and its orientation ($\Delta \theta$ $=0.1^o$). Moreover, electric field-dependent fringe visibility is observed, which can be tuned by $E_0$ and $\theta$. The potential barriers arising from a spatially modulated electric field are observed to cause significant distortions in the Talbot patterns when the modulation length ($\lambda'$) is equal to the de Broglie wavelength ($\lambda_{dB}$). Sidebands are observed in the Talbot pattern due to the efficient transfer of energy from the oscillating field to the wave packet when the frequency of oscillation ($\omega$) is of the order of $\omega_0$ ($=2\pi/T_0$), where $T_0$ is the interaction time. This study will be helpful in uniform electric field-controlled precision metrology, developing a highly sensitive electric field sensor based on Talbot interference, and precisely aligning the matter wave optical setup. Furthermore, the sidebands in the Talbot fringe can be used as a precise tool as momentum splitter in matter wave interferometry.
	
\end{abstract}


\maketitle
\section{\label{introduction}Introduction}

In near-field diffraction of a plane wave, the diffraction grating creates self-images at regular distances away from it. This phenomenon was observed in classical optics in 1836 by Henry Fox Talbot \cite{doi:10.1080/14786443608649032} and is known as the Talbot effect \cite{PhysRevLett.104.183901, PhysRevLett.118.133903, hall2021space, PhysRevLett.100.030202, Rodrigues:17}. Since the discovery, the Talbot effect has been used in many optical applications ranging from surface profiling of transparent objects \cite{Thakur:05} and testing of beam collimation \cite{Kothiyal:87} to several precision measurements, such as vibration in real-time \cite{PRAKASH2000251}, periodic phase modulation \cite{Hamam:95}, aberration \cite{Takeda:84}, focal dimension of lenses \cite{Nakano:85}, small tilt angle \cite{Nakano:87}, grating parameters \cite{Photia:19}, and temperature profile of gaseous flame \cite{Shakher:94}. Recently, this effect was used to detect optical vortices \cite{Panthong:18} and the construction of high-resolution spectrometers \cite{Han:19}.

Due to the wave nature, quantum particles also show diffraction effects, which are similar to those in the case of light. After the first observation of wave nature for electron \cite{doi:10.1073/pnas.14.4.317}, several experiments were performed on matter wave diffraction to study the quantum effects \cite{PhysRev.90.490, PhysRevLett.75.2633, PhysRevLett.68.472, PhysRevLett.56.827, PhysRevLett.61.1580, Gronniger_2006}, which resulted in the development of matter wave interferometers. Similar to classical optics, the Talbot effect was observed experimentally in matter wave optics for electron \cite{PhysRevB.93.104305}, positron \cite{doi:10.1126/sciadv.aav7610}, neutral atoms \cite{Nowak:97, PhysRevA.51.R14, Mark_2011, Muller_2020}, and complex molecules \cite{Kohstall_2011, hackermuller2007optical, PhysRevA.76.013607, gerlich2007kapitza, PhysRevLett.88.100404,Eibenberger_2011, bateman2014near, eibenberger2013matter}.

Recently, there has been a huge surge in research on quantum technologies for computing, materials, communication, and instrumentation. These efforts drive the formulation of novel concepts and the development of quantum devices, such as sensors and circuitry for quantum memories. In this regard, the near-field diffraction of matter waves has become an important tool for studying quantum effects and measurement of fundamental quantities. Therefore, there is a need to understand the phenomenon under different conditions. Several theoretical as well as experimental investigations have been carried out in this direction. The Talbot effect in the near-field diffraction was studied by solving Fresnel-Kirchhoff integral in paraxial approximation for a three-grating setup with uncollimated incident particles \cite{B_Brezger_2003}. Interference pattern formation in Talbot-Lau interferometers was studied by combining semiclassical scattering theory with a phase space formulation accounting for arbitrary grating interactions and realistic beam properties \cite{PhysRevA.78.023612}. The same phenomenon has been studied for large molecules by considering the reconstruction of the Wigner distribution function (WDF), where the experimental conditions of the WDF reconstruction have been estimated for the Talbot interference fringes \cite{Lee_2012}. The diffraction of cold Rb atoms by a single-, double-, and multiple-slit grating were studied in the near-field to the far-field region, including the transition region, using the Feynman path integration method \cite{Srisuphaphon_2019, Artyotha_2018}. Another theoretical study discussed the possibility of Talbot self-imaging for relativistic matter waves \cite{saif2012talbot}.

Apart from these theoretical studies, numerous experimental studies have been performed in parallel. Three equally spaced transmission gratings were employed to study the transition from the classical description of the diffraction pattern with Xe atoms to the Talbot-Lau interference in the quantum regime with H atoms \cite{Muller_2020}. Furthermore, the feasibility of interferometers based on charged particles, such as protons and antiprotons, have been discussed. The matter wave interference of positrons in a Talbot-Lau interferometer was demonstrated \cite{doi:10.1126/sciadv.aav7610}. Experiments have been performed on the near-field diffraction for atoms, such as He \cite{Nowak:97}, Na \cite{ PhysRevA.51.R14}, and Cs \cite{Mark_2011}. Scalar polarizabilities of complex molecules, such as C\textsubscript{60}, C\textsubscript{70} \cite{hackermuller2007optical, PhysRevA.76.013607, PhysRevLett.88.100404}, C\textsubscript{30}H\textsubscript{12}F\textsubscript{30}N\textsubscript{2}O\textsubscript{4} \cite{gerlich2007kapitza}, C\textsubscript{49}H\textsubscript{16}F\textsubscript{52} \cite{Eibenberger_2011} and silicon nanoparticle \cite{bateman2014near}, were measured using the near-field matter wave interferometry.

Numerical studies have also been performed to study ultrafast revivals of quantum states associated with the temporal Talbot effect, which results in higher-order harmonic generation and suggests developing subnanometer ultrafast Talbot interferometers \cite{G_Cabrera_2021}. Kazemi et al. have performed numerical investigations on the effect of decoherence on Talbot images \cite{Kazemi_2013}. Studies on near-field diffraction of charged or neutral antimatter particles suggest that the van der Waals interactions for the polarizable particle are critical to the fringe visibility \cite{Sala_2015}.

Studies so far suggest that the Talbot effect in near-field diffraction is not only an interesting matter wave phenomenon but also leads to the various applications mentioned above. However until now, there are no studies on the effect of external electric fields on the near-field diffraction of charged particles, such as protons, and how the electric fields control the Talbot pattern, including their potential applications.

In this study, near-field diffraction of protons by a nano-structured metallic grating in the presence of (a) uniform, (b) spatially modulated, and (c) temporally modulated electric fields are investigated. Time-domain simulations are performed using the generalized finite difference time domain (GFDTD-Q) method for quantum systems to simulate two-dimensional Gaussian wave packets of protons. The effect of the applied uniform electric field strength ($E_0$) and its orientation ($\theta$) on the Talbot pattern are investigated. The diffracted wave packet is analyzed in the wave number space ($k_x$, $k_y$) to investigate the quantization of axial ($k_x$) and lateral ($k_y$) wave numbers. The results show that the fringe pattern gets shifted in the transverse direction. Moreover, it is found that $E_0$ and $\theta$ can tune the fringe shift and visibility. Distortions are observed in the Talbot fringe in the case of spatially modulated electric field when the modulation length ($\lambda'$) is equal to the de Broglie wavelength ($\lambda_{dB}$). Due to the efficient energy transfer from the oscillating field to the wave packet, multiple sidebands are observed for the first time in the Talbot pattern when the oscillation frequency ($\omega$) is of the order of $\omega_0$ ($=2\pi/T_0$) with $T_0$ being the interaction time. Being a lighter ion, a proton responds easily, even to a small electric field. The findings of this research have important implications in electric field-controlled precision metrology, to develop a highly sensitive electric field sensor based on the Talbot effect and in precise alignment of matter wave optical setup. Furthermore, the sidebands observed in the Talbot fringe will be useful as momentum splitters in matter wave interferometry. 

The article is structured as follows. Section \ref{methods} describes the numerical method for time-domain simulation of matter waves. In Sec. \ref{result_and_discussions}, wave packet simulation results are presented and discussed. Finally, conclusions are made in Sec. \ref{conclusions}.
\begin{figure}
	\includegraphics[scale=0.96]{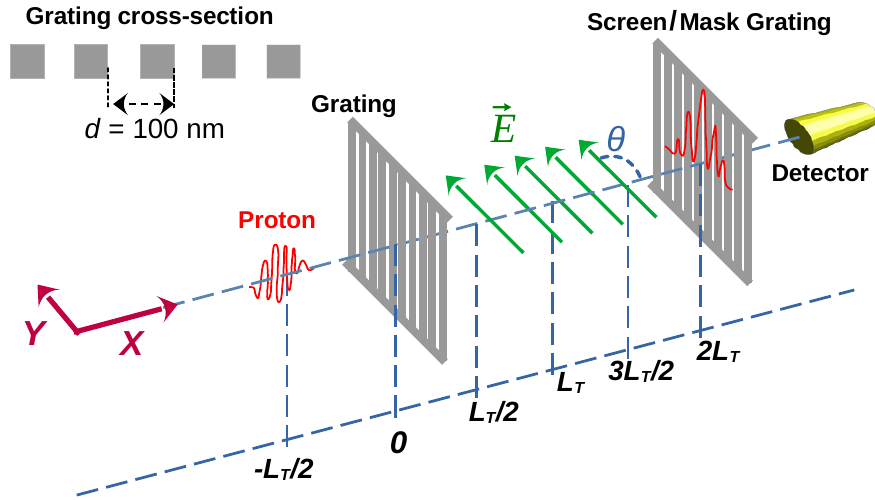}
	\caption{Schematic diagram of the diffraction setup under investigation. A metallic grating having periodicity $d=100$ nm and opening fraction $f=50\%$ is used to diffract protons. The incident particles move along the $+x$ direction with an initial velocity $v=394$ m/s. Similar to the diffraction grating, a mask grating is placed at $x=2L_T$, where $L_T (=d^2/\lambda_{dB}$) is the Talbot length with $\lambda_{dB}$ being the de Broglie wavelength of protons. An electric field ($\vec{E}$) is applied in the region $L_T/2\leq x \leq 3L_T/2$ behind the diffraction grating. The electric field $\vec{E}$ makes an angle $\theta$ with the $x$-axis. A detector is placed behind the mask grating to detect the diffracted particles at various positions along the $y$-axis.}
	\label{diffraction_setup}
\end{figure}

\begin{figure}
	\includegraphics[scale=0.95]{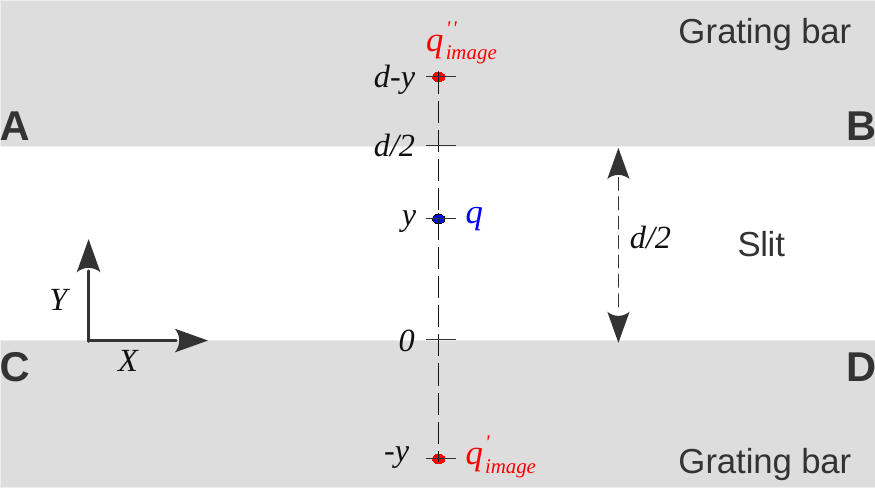}
	\caption{A cross-sectional view of the diffraction grating. ABCD is one of the slit openings in the grating. AB and CD are the planes of two neighboring bars of the grating. Since the grating is made of metal, image charges ($q^{'}_{image}$ and $q^{''}_{image}$) are induced due to a charge $q$ situated at a distance $y$ from the lower plane (CD). The width of the slit is $d/2$.}
	\label{image_charge_plot}
\end{figure}
\section{Method}
\label{methods}
The setup for Talbot-Lau interferometer consists of three gratings, where the first grating is used to make a coherent source of matter waves \cite{PhysRevA.81.031604, Juffmann_2013}. The diffraction of coherent waves occurs in the second grating, while the third grating acts as a mask grating. Thus, self-imaging by the second grating can be considered for a complete description of the diffraction phenomenon in the interferometer \cite{B_Brezger_2003}. The setup under consideration consists of a metallic grating having periodicity $d=100$ nm and opening function $f=50\%$, as shown in Fig. \ref{diffraction_setup}. The first grating is absent as we consider incident monoenergetic and coherent protons that move along the $x$-direction with a velocity $\text{v}=394$ m/s. The corresponding de Broglie wavelength $\lambda_{dB}=1$ nm and kinetic energy $E_{k0}=0.813$ meV. The ion beams are extracted from a cryo-plasma \cite{Stauss_2013, noma2011electron}, and then, the beams are collimated to the required energy, in an experiment that is currently under development in our laboratory. Usually, these experiments are performed in high vacuum ($<10^{-7}$ mbar) to minimize dissipation through ion-atoms collisions.  The image of the diffraction grating repeats at the Talbot length $L_T=d^2/\lambda_{dB}$ \cite{doi:10.1080/14786448108626995}. An electric field ($\vec{E}$) is applied in the region $L_T/2\leq x \leq 3L_T/2$, as shown in Fig. \ref{diffraction_setup}. The diffraction phenomena of protons can be investigated by solving the two-dimensional ($x$, $y$) time-dependent Schrödinger's equation given by
\begin{align}
	i\hbar \frac{\partial \psi (x,y,t)}{\partial t} &= \bigg[- \frac{\hbar^2}{2m} \nabla^2 
	+ V_g(x,y) + eV_{image}(x,y) \nonumber \\&+ e V_E(x,y,t) \bigg] \psi(x,y,t),
	\label{sch_eq}
\end{align}
where $i=\sqrt{-1}$, and $\psi(x,y,t)$, $m$, $e$, and $\hbar$ are the wavefunction, mass of proton, electronic charge, and reduced Planck's constant, respectively. $V_g(x,y)$ is the geometric potential considering the grating bars as hard walls \cite{ghalandari2020fractional,barman2022time}. $V_{image} (x,y)$ is the electrostatic potential due to the induced image charge in the conducting bars of the grating, and  $V_E(x,y,t)$ is the electrostatic potential due to the applied electric field $\vec{E}$.

The geometric potential of a periodic grating can be expressed as \cite{PhysRevA.101.022506, doi:10.1021/acs.jpclett.6b02991},
\begin{equation}
	V_g(x,y) = V_{g0}
	\begin{cases}
		\sum_{m=-\infty}^{+\infty} \prod \bigg( \frac{y-md}{d/2}\bigg), & \text{ if $0\leq x\leq l$ and all $y$} \\
		0, & \text{if otherwise}\\

	\end{cases}      
	\label{unit_impulse_function} 
\end{equation}
where $l$ and $V_{g0}$ are the grating thickness and height of the geometrical potential barrier, respectively. The cross-section of the grating bars are assumed to be rectangular, as shown in Fig. \ref{diffraction_setup}. The value of $V_{g0}$ is taken as $\sim 30E_{k0}$, which is well above the incident particle energy, to minimize the effect of quantum tunneling through the grating bars \cite{PhysRevA.78.023612, Castellanos-Jaramillo_2018}. The $\prod (u)$ function is the unit impulse function, which is given by,
\begin{equation}
	\prod(u) =
	\begin{cases}
		0, & \text{if $|u|> 1/2$} \\
		1, & \text{if $|u|< 1/2$} \\
		\frac{1}{2}, & \text{if $|u|= 1/2$} \\
	\end{cases}      
	\label{unit_impulse_function} 
\end{equation}

When a proton passes through the slits, it interacts with the walls of the grating bars. The proton-wall interaction can be described by the image charge potential, which can be obtained from the well-known image charge solution of Laplace's equation by treating the slit walls as infinite planes. This method has also been used to evaluate particle-grating interactions in previous studies of electron diffraction \cite{barwick2006measurement}. Considering the grating walls as infinite planes, as shown in Fig. \ref{image_charge_plot}, the image charge potential is obtained as \cite{barwick2006measurement}
\begin{align}
V_{image}(x,y)=\frac{e}{8\pi \epsilon_0} \bigg[\frac{q^{'}_{image}}{y} + \frac{q^{''}_{image}}{(d/2-y)}  \bigg], \text{ if $0< y<d/2$} \nonumber \\ \text{and $0\leq x \leq l$}
\end{align}
where $\epsilon_0$ is the electric permittivity of free space. For metal grating, the induced image charges can be obtained as $q^{'}_{image}=q^{''}_{image}=-e$.

Before starting the numerical calculations, Eq. \textcolor{blue}{(\ref{sch_eq})} is nondimensionalized in arbitrary units of space and time to make it computationally advantageous. The spatial and temporal domains are transformed as  $\vec{r}(x,y)=\gamma \vec{R}(X,Y) $ and $t=\tau T$, respectively. Where $\gamma$ and $\tau$ have the dimension of length (m) and time (s), respectively. Therefore, the 2D Laplacian in the new coordinate system can be expressed as $\nabla^2 = (1/\gamma^2) \nabla_{\gamma}^2 =(1/\gamma^2)(\partial ^2/\partial X^2 + \partial ^2/\partial Y^2) $. Similarly, the potential energies are scaled as $\phi_{g} (X,Y)=(1/V_0)V_g(x,y)$, $\phi_{image} (X,Y)=(e/V_0)V_{image}(x,y)$, and $\phi_{E} (X,Y,T)=(e/V_0)V_E(x,y,t)$ with the scaling factor $V_0=\hbar^2/2m\gamma^2$, which has a dimension of energy. The functions $\phi_{g}(X,Y)$, $\phi_{image}(X,Y)$, and $\phi_{E}(X,Y,T)$ are dimensionless functions. Thereafter, the nondimensionalized form of Eq. \textcolor{blue}{(\ref{sch_eq})} is obtained by taking $\gamma=\sqrt{ \hbar^2/2mV_0}$ and $\tau=2m\gamma^2/\hbar$, as
\begin{align}
	i\frac{\partial \psi (R,T)}{\partial T} &= \bigg[- \nabla_{\gamma}^2 + \phi_{g}(R) +\phi_{image}(R)  + \phi_E(R,T)\bigg] \psi(R,T),
	\label{sch_eq6}
\end{align}
where the wavefunction $\psi(R,T)$ can be separated into real ($\psi_{real} (R,T)$) and imaginary ($\psi_{imag} (R,T)$) parts  as $\psi(R,T) = \psi_{real} (R,T) + i\psi_{imag} (R,T)$. 
\begin{figure}
	\includegraphics[scale=0.93]{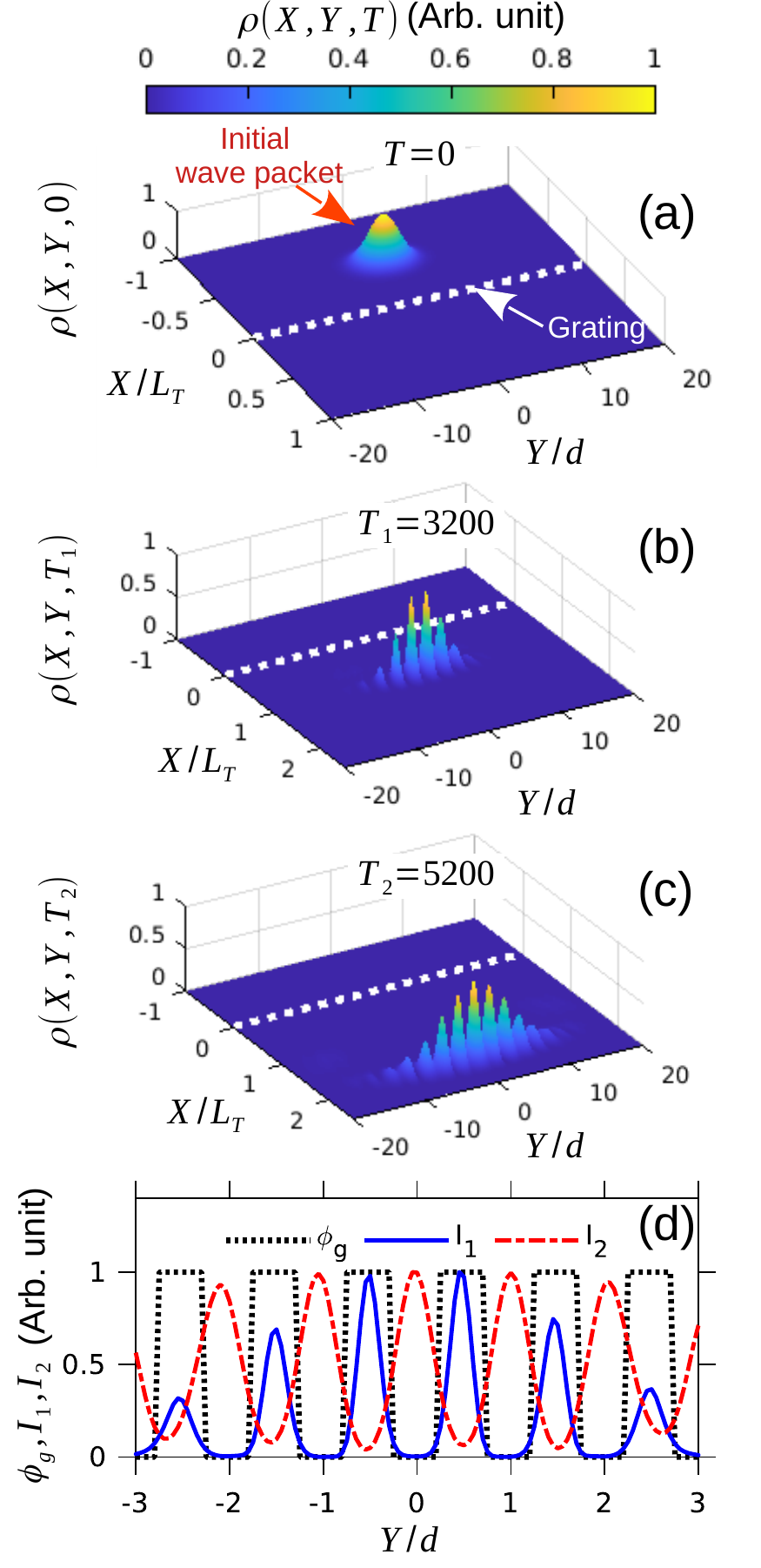}
	\caption{(a) Distribution of the probability density ($\rho(X, Y, 0)=|\psi(X, Y, 0)|^2$) of the initial Gaussian wave packet moving along the $+x$ direction. Distribution of the probability density ($\rho(X, Y, T)=|\psi(X, Y, T)|^2$) of the diffracted wave packet at (b) $T_1=3200$ and (c) $T_2=5200$, when the wave packet reached at $X_1=L_T$ and $X_2=2L_T$, respectively. (d) Comparison of the intensities in the diffraction patterns, $I_1=|\psi(L_T,Y,T_1)|^2$ and $I_2=|\psi(2L_T,Y,T_2)|^2$ with the grating function ($\phi_g (0,Y)$).}
	\label{talbot_effect_plot}
\end{figure}

The real and imaginary parts of Eq. \textcolor{blue}{(\ref{sch_eq6})} are equated, and the following coupled equations are obtained,
\begin{align}
	\frac{\partial \psi_{real} (R,T)}{\partial T} = \bigg[- \nabla_{\gamma}^2 + \phi_{g}(R) &+\phi_{image}(R) \nonumber \\ &+ \phi_E(R,T)\bigg] \psi_{imag}(R,T),
	\label{psi_r}
\end{align}
and
\begin{align}
	\frac{\partial \psi_{imag} (R,T)}{\partial T} = \bigg[\nabla_{\gamma}^2 -  \phi_{g}(R) &-\phi_{image}(R)  \nonumber \\ & - \phi_E(R,T)\bigg] \psi_{real}(R,T).
	\label{psi_i}
\end{align}

The coupled Eqs. \textcolor{blue}{(\ref{psi_r})} and \textcolor{blue}{(\ref{psi_i})} are solved using generalized finite difference time domain (GFDTD-Q) method \cite{MOXLEY20122434} for quantum mechanics to obtain the time evolution of an initial wave function $\psi(X, Y, T=0$). The time evolution of $\psi_{real} (X,Y,T)$ and $\psi_{imag} (X,Y,T)$ is obtained using the leapfrog method.

The GFDTD-Q method is a powerful numerical tool for an explicit solution of linear and nonlinear time-dependent Schrödinger's equations, which appear in various physical systems \cite{MOXLEY20131834, DECLEER2022113881, WILSON2019279, MOXLEY2015303}. In this method, the temporal and spatial differential operators are approximated as centered difference equations. The scheme is stable and accurate up to $(\Delta T)^3$ and $(\Delta X)^5$. A complete study on the stability analysis of this method may be found in Ref. \cite{MOXLEY20122434, DECLEER2022113881}. Physical systems having potentials of any arbitrary configuration can be simulated using the GFDTD-Q method \cite{Wayan_Sudiarta_2007,soriano2004analysis, sullivan2002determination, duan2021computational, DECLEER2021113023, GHAFOURI2020103502, Talebi_2021, doi:10.1080/23746149.2018.1499438, Talebi_2016, sullivan2004time}. Moreover, this method has already been applied in previous studies to simulate matter wave diffraction phenomena \cite{Castellanos-Jaramillo_2018, barman2022time}. A detailed formulation of the GFDTD-Q method and the numerical procedures can be found in Ref. \cite{barman2022time}.
\begin{figure*}
	\includegraphics[scale=1.0]{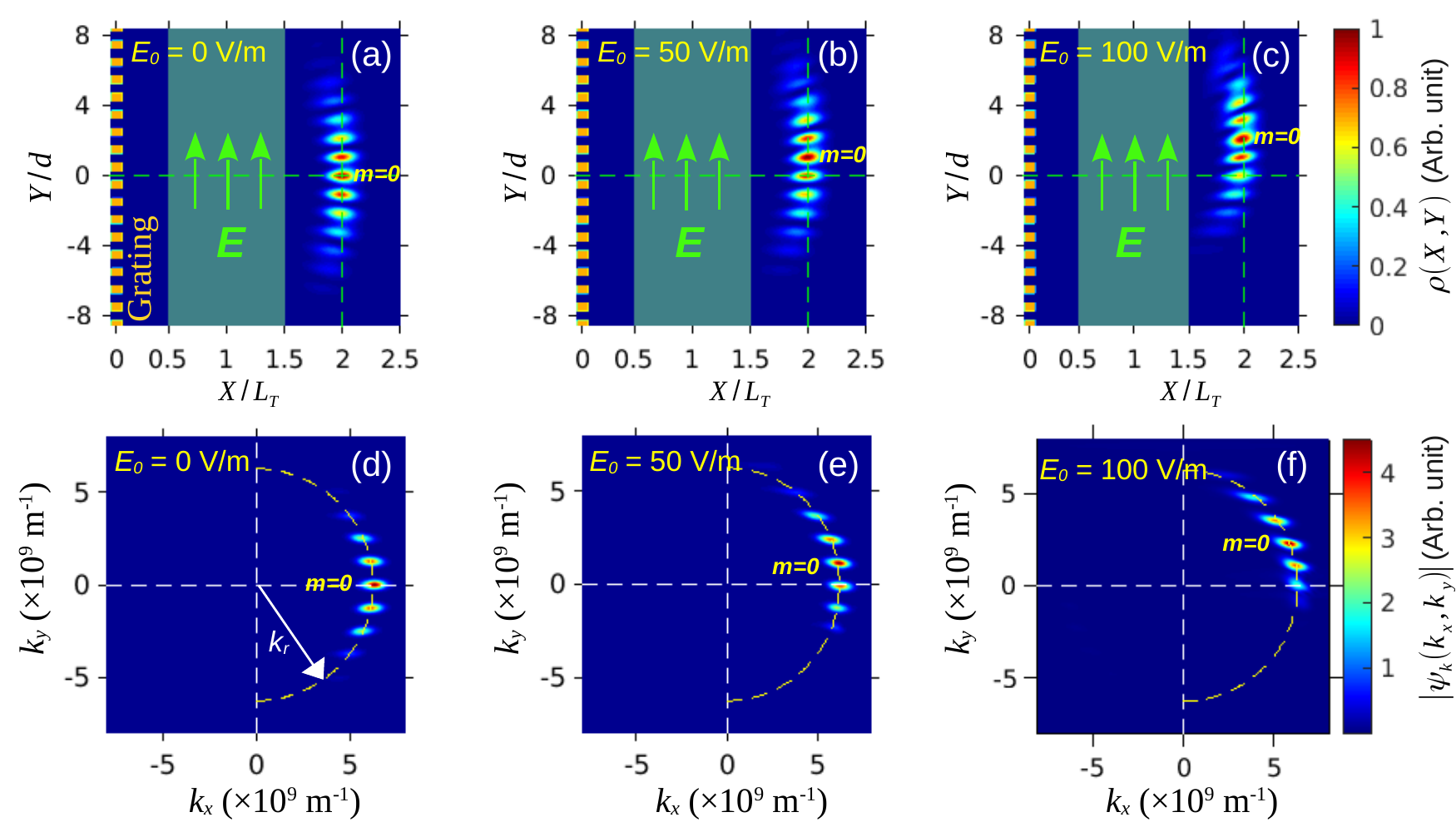}
	\caption{Final shape of the diffracted wave packet at $T_2=5200$: the panels (a)-(c) are shown in position space ($X, Y$), and panels (d)-(f) are shown in wave number space ($k_x$, $k_y$). The panels (a) and (d) are plotted for the magnitude of the applied electric field $E_0=0$ V/m; (b) and (e) are plotted at $E_0=50$ V/m; (c) and (f) are plotted at $E_0=100$ V/m. In all the cases, the electric field is applied in the transverse direction ($\theta=90^o$) to the direction of incidence ($x$-axis) of the particle.}
	\label{probability_distribution_xy_kxky_plot}
\end{figure*}

The initial wave packet is taken as a two-dimensional Gaussian wave packet given by
\begin{align}
	\psi(X,Y,T=0)&= C \exp \bigg[-\frac{1}{2} \bigg\{\frac{(X-X_0)^2}{\sigma_X^2}  + \frac{(Y-Y_0)^2}{\sigma_Y^2}\bigg\}  \nonumber \\
	& + ik(X-X_0) \bigg],
\end{align}
where $C$ is the normalization constant and ($X_0, Y_0$) is the center of the wave packet at $T=0$. $\sigma_X$ and $\sigma_Y$ are the width of the wave packet along $X$ and $Y$-axes, respectively. $k$ ($=2\pi/\lambda_{dB}$) is the wave number of the wave packet. 

A 2D region of $501\times 501$ square grids ($\Delta X=\Delta Y$) has been considered in the simulation. The spatial and temporal scaling factors are taken as $\gamma=2.5$ {\AA} and $\tau=1.982 \times 10^{-12}$ s. The details of the simulation parameters are listed in Table \ref{simulation_parameters}. The convergence in the simulation is confirmed by checking the unit norm of the initial wave packet $\psi(X, Y, T=0)$ at each time step.

In the simulations, artificial reflections occur when the wave packet hits the boundary of the numerical box. Hence, absorptive boundary conditions are employed to avoid such unwanted reflections from the boundaries. At each time step, the wave function is multiplied by a damping function $\zeta(X)$ given by \cite{Herwerth_2013}
\begin{align}
	\zeta(X) =\frac{1}{1+\exp(-\frac{X-X_0}{\Lambda})},
	\label{damping_func}
\end{align}
where $\Lambda$ ($=5$) and $X_0$ ($=0$) are two parameters that determine the sharpness and position of the damping function for the left boundary. The $\zeta(X)$ function has a unit value for all $X$ except near the boundary where the wave function needs to be suppressed to minimize artificial reflections. The corresponding damping function for the right boundary is obtained by choosing the proper values of $X_0$ ($=X_{max}$) and $\Lambda$ ($=-5$). Similar damping functions have been implemented in the simulation for the upper and lower boundaries.

The wave packet in the wave number space ($k_x$, $k_y$) can be obtained by Fourier transformation as
\begin{align}
	\psi_k(k_x, k_y,t)=\frac{1}{2\pi}\int \int \psi(x,y,t)e^{-i(k_x x+k_y y)}dx dy.
	\label{fourier_transformation}
\end{align}
\begin{figure}
	\includegraphics[scale=1.0]{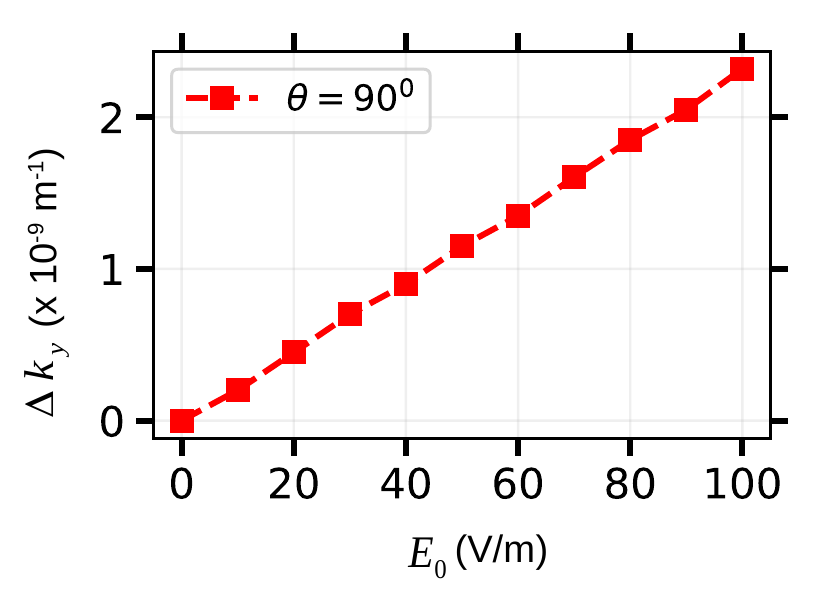}
	\caption{Variation of the change in transverse wave number $\Delta k_y$ of the diffracted wave packet with the applied uniform electric field strength $E_0$. The values of $\Delta k_y$ are obtained at $X_2=2L_T$ for the orientation of the electric field $\theta=90^o$.}
	\label{change_in_ky}
\end{figure}
\begin{figure}
	\includegraphics[scale=0.95]{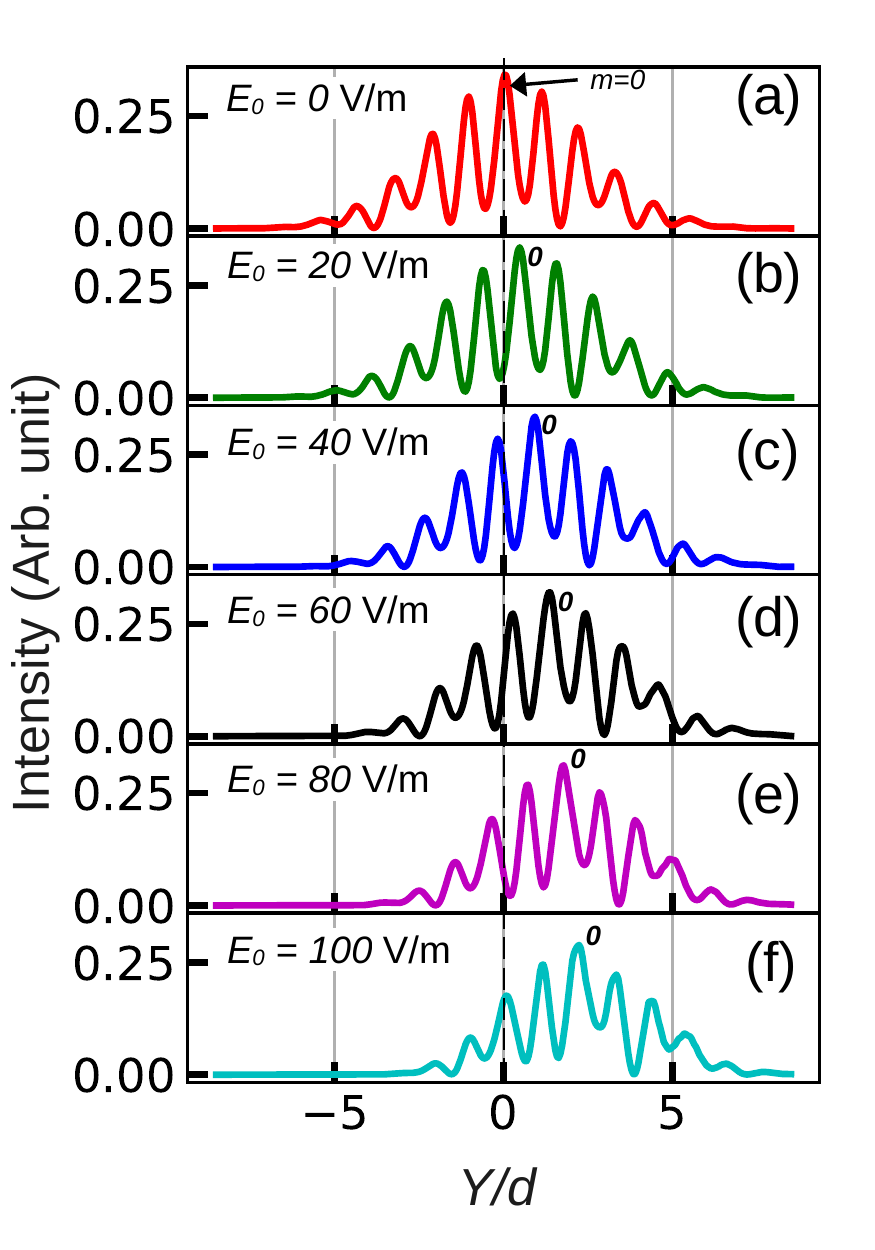}
	\caption{Fringe pattern at $X_2=2L_T$ for different magnitudes ($E_0$) of the applied electric fields in the transverse direction ($\theta=90^o$). (a) $E_0=0$ V/m, (b) $E_0=20$ V/m, (c) $E_0=40$ V/m, (d) $E_0=60$ V/m, (e) $E_0=80$ V/m, and (f) $E_0=100$ V/m.}
	\label{fringe_pattern_xy_plot}
\end{figure}
\begin{figure}
	\includegraphics[scale=0.95]{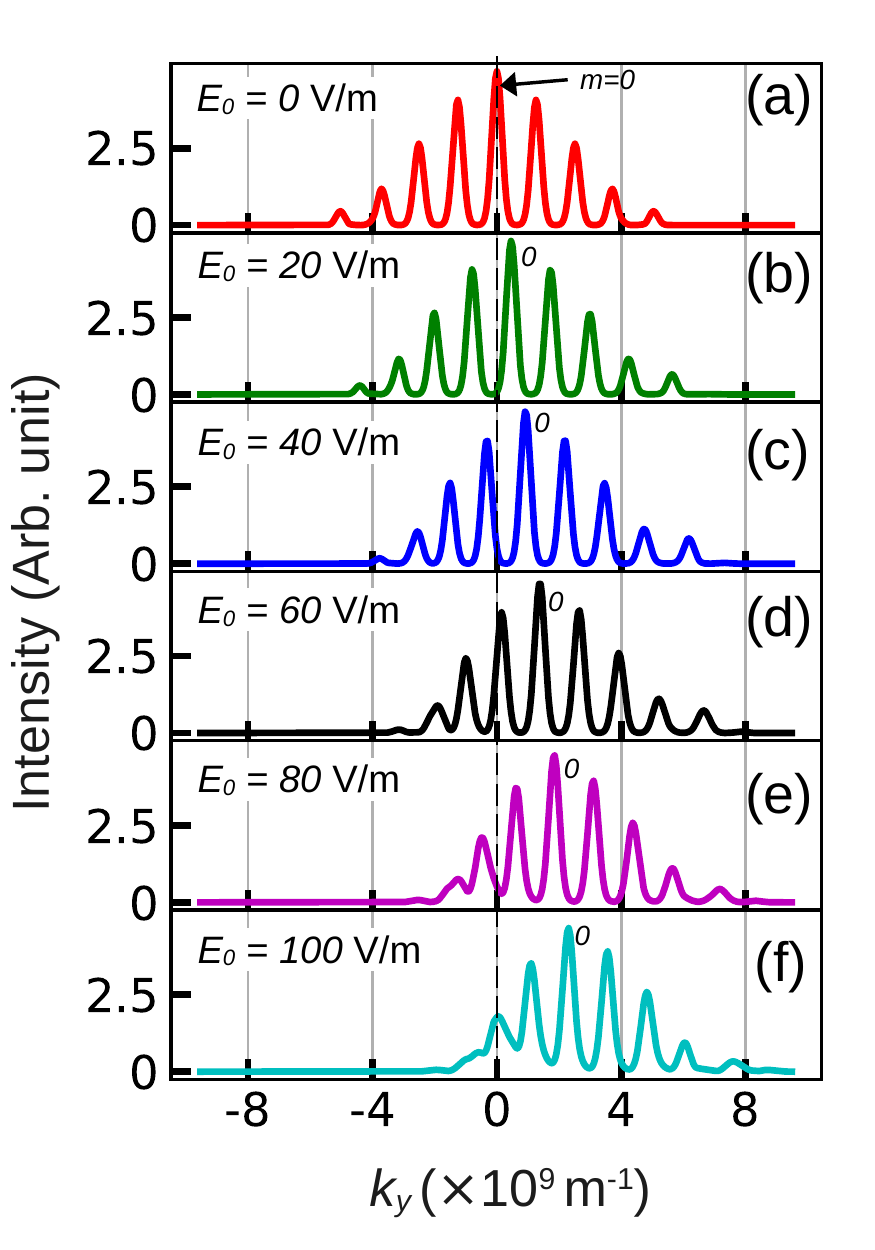}
	\caption{Fringe pattern in the momentum space ($k_X, k_y$) for different magnitudes of electric fields (a) $E_0=0$ V/m, (b) $E_0=20$ V/m, (c) $E_0=40$ V/m, (d) $E_0=60$ V/m, (e) $E_0=80$ V/m, and (f) $E_0=100$ V/m. The electric field is applied in the transverse direction ($\theta=90^o$).}
	\label{fringe_pattern_kxky_plot}
\end{figure}
\begin{table}[ht]
	\centering
	\caption{Values of the parameters used in the simulation.}
	\label{simulation_parameters}
	\begin{tabular}[t]{lccc}
		\hline \hline
		Parameters & Values\\
		\hline
		$\sigma_X$ & 50 nm\\
		$\sigma_Y$ & 250 nm\\
		($X_0$, $Y_0$) & ($-L_T/2$, $0$)\\
		de Broglie Wavelength $(\lambda_{dB})$ &  1 nm\\
		Velocity of the incident proton ($\text{v}$) & $394$ m/s\\
		Periodicity of the grating ($d$) & 100 nm \\
		Opening function of the grating ($f$) & $50\%$\\
		Thickness of the grating ($l$) & 50 nm\\
		Geometrical potential barrier height ($V_{g0}$) & 24.390 meV \\
		$V_0$ & 0.332 meV\\
		$\gamma$ & 2.5 {\AA}\\
		$\tau$ & $1.982\times 10^{-12}$ s\\
		Magnitude of the electric field ($E_{0}$) & $0-100$ V/m \\
		Orientation of the electric field ($\theta$) & $0^o - 90^o$ \\
		
		\hline\hline
	\end{tabular}
\end{table}
\section{Results and discussions} \label{result_and_discussions}
\subsection{Near-field diffraction without electric field}
Fig. \ref{talbot_effect_plot} \textcolor{blue}{(a)} shows the distribution of the initial wave packet $\rho(X,Y,T=0)$. The wave packet evolves with time and gets diffracted by the grating. The distribution of the probability density of the wave packet behind the diffraction grating at time $T_1=3200$ and $T_2=5200$ are shown in Figs. \ref{talbot_effect_plot} \textcolor{blue}{(b)} and \textcolor{blue}{(c)}, respectively. It can be noticed that at $T_1$ and $T_2$, the diffracted wave packet reached at $X_1=L_T$ and $X_2=2L_T$, respectively. At later times ($T_2$), higher order peaks ($m=\pm6$) are observed (see Fig. \ref{talbot_effect_plot} \textcolor{blue}{(c)}), which are not observed at $T_1$ (maximum order $m=\pm4$) (see Fig. \ref{talbot_effect_plot} \textcolor{blue}{(b)}). This happens because in the near-field regime, the shape of the diffracted wave packet evolves with time, and there exists a time delay between the central maxima and the higher order peaks, as reported in an earlier investigation \cite{barman2022time}. Furthermore, from the distribution of diffracted wave packets at $T_1=3200$ and $T_2=5200$, it is clear that the fringe periodicity becomes equal to the grating periodicity ($d$). This phenomenon is known as the Talbot effect, which is distinct from the far-field diffraction, where the probability density peaks at an angle $\alpha=\sin^{-1} (m \lambda_{dB}/d)$ behind the diffraction grating \cite{RevModPhys.84.157}.

For further understanding of the diffraction properties, the intensities of the fringe pattern on a screen placed at $X_1=L_T$ and $X_2=2L_T$ are obtained as $I_1=|\psi(L_T, Y, T_1)|^2$ and $I_2=|\psi(2L_T, Y, T_2)|^2$, respectively. The variation of $I_1$, $I_2$, and the grating function $\phi_g(0,Y)$ along the transverse direction ($Y$) are shown Fig. \ref{talbot_effect_plot} \textcolor{blue}{(d)}. The figure shows that both $I_1$ and $I_2$ vary sinusoidally with a decaying amplitude on either side from the center ($Y=0$) of the grating. Moreover, $I_1$ varies with $180^o$ out of phase to the grating function, whereas $I_2$ varies with the same phase as $\phi_g(0, Y)$. These characteristics of Talbot patterns, which are observed in earlier studies as well \cite{RevModPhys.84.157}, confirm the accuracy of the simulation techniques used in this study.

\subsection{Near-field diffraction in a uniform electric field}
In this section, we have investigated the time evolution of the diffracted wave packet in the near-field region under the influence of uniform electric fields given by

\begin{align}
 \vec{E} (x,y)= E_0(\hat{x} \cos{\theta} + \hat{y} \sin{\theta}),
\end{align}
where $E_0$ and $\theta$ are the magnitude of $\vec{E}$ and the angle made by $\vec{E}$ with $x$-axis, respectively. The electric field is employed in the region $L_T/2\leq x \leq 3L_T/2$.

The corresponding distribution of the electrostatic potential is obtained as
\begin{align}
	V_{E}(x,y)= 
	\begin{cases}
		-E_0 (x\cos\theta + y \sin\theta),& \text{if $L_T/2 \leq x \leq 3L_T/2$}  \nonumber \\ &\text{and all $y$} \\
		0, & \text{if otherwise}\\
	\end{cases}   
\end{align}

The effect of the magnitude ($E_0$) and orientation ($\theta$) of the applied electric field $\vec {E}$ on the diffraction patterns, fringe shift, and visibility are investigated. Further, the wave packets are analyzed in the wave number space ($k_x$, $k_y$) to investigate its transverse quantization after diffraction.
\subsubsection{Diffraction pattern}
Fig. \ref{probability_distribution_xy_kxky_plot}\textcolor{blue}{(a)}-\textcolor{blue}{(c)} shows the distribution of $\rho(X,Y,T)$ of the diffracted wave packet at time $T_2=5200$ for different values of $E_0$ ($=0$, $50$ and $100$ V/m). In all the cases, the electric fields are applied in the transverse direction ($\theta=90^o$). A well-formed pattern is observed for $E_0=0$ V/m, as shown in Fig. \ref{probability_distribution_xy_kxky_plot}\textcolor{blue}{(a)}. Figs. \ref{probability_distribution_xy_kxky_plot}\textcolor{blue}{(b)} and \textcolor{blue}{(c)} show that the fringe patterns get shifted along $+y$-direction and the amount of fringe shift increases with increase in $E_0$. The fringe shifts are calculated to be, $s=1.1 \times d$ and  $2.15 \times d$ ($d=100$ nm) for $E_0=50$ and $100$ V/m, respectively. Moreover, for large electric fields, higher order peaks ($m\geq+3$) get overlapped with each other, as observed in Fig. \ref{probability_distribution_xy_kxky_plot}\textcolor{blue}{(c)}.

Next, the diffracted wave packets are analyzed in the wave number space ($k_x$, $k_y$), which are obtained using Eq. \textcolor{blue}{(\ref{fourier_transformation})}. Figs. \ref{probability_distribution_xy_kxky_plot}\textcolor{blue}{(d)}-\textcolor{blue}{(f)} show the distribution of $|\psi_k(k_x,k_y,T)|$ for different values of $E_0$ ($=0$, $50$ and $100$ V/m) at $\theta=90^o$ and $T_2=5200$. A semi-circular distribution having radius of $k_r=\sqrt{k_x^2 + k_y^2}$ is observed among the the diffraction peaks in $|\psi(k_x,k_y,T)|$. It is seen that the value of $k_r$ is equal to the wave number of the initial wave packet, $k_0=2\pi/\lambda_{dB}=6.28 \times 10^{-9}$ m\textsuperscript{-1}. More interestingly, it is observed that $k_r$ remains constant irrespective of $E_0$ because the electric field is applied in a direction perpendicular to the motion of the particle. Therefore, no work is done by $\vec E$ on the particle, and the momentum remains conserved. However, the diffracted wave packet gets deflected in the perpendicular direction ($+y$-axis). Hence, the transverse wave number $k_y$ increases with the increase in $E_0$ while keeping $k_r$ a constant. The change in transverse wave number ($\Delta k_y$) with $E_o$ for $\theta = 90^o$ is shown in Fig. \ref{change_in_ky}.

The fringe patterns on a screen placed at $X_2=2L_T$  are shown in Figs. \ref{fringe_pattern_xy_plot}\textcolor{blue}{(a)}-\textcolor{blue}{(f)} for different values of $E_0$ when the field is applied in the transverse direction ($\theta=90^o$). From Figs. \ref{fringe_pattern_xy_plot}\textcolor{blue}{(a)}-\textcolor{blue}{(f)}, it is clear that the periodicity in the fringe patterns, which is same as the grating periodicity $d$, remains constant in all the cases. Moreover, all the peaks in the diffraction pattern get shifted equally, and this shift increases with the increase in $E_0$. Further, it can be noticed that at large electric fields ($E_0\geq 80$ V/m), the visibility of the higher order diffraction peaks ($m\geq +3$) get reduced, as shown in Fig. \ref{fringe_pattern_xy_plot}\textcolor{blue}{(f)}. A similar trend is observed in the diffraction pattern obtained in the momentum space as shown in Fig. \ref{fringe_pattern_kxky_plot}\textcolor{blue}{(a)}-\textcolor{blue}{(f)}. The amplitude of the higher order peaks ($m\geq +3$) decreases at large electric fields (see Fig. \ref{fringe_pattern_kxky_plot}\textcolor{blue}{(f)}), which is also clear from Fig. \ref{probability_distribution_xy_kxky_plot}\textcolor{blue}{(f)}.

\begin{figure}
	\includegraphics[scale=1.0]{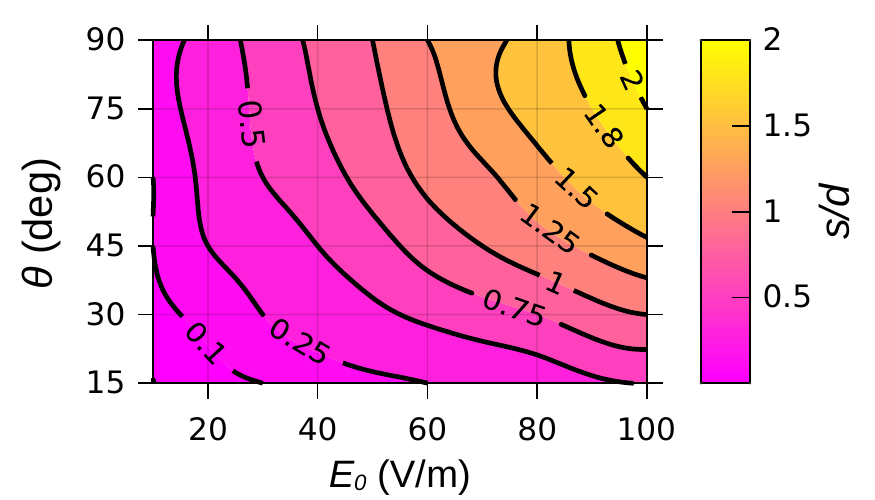}
	\caption{Variation of fringe shift normalized by the grating period ($s/d$) with the magnitude $E_0$ and orientation ($\theta$) of the applied uniform electric field.}
	\label{fringe_shift_colormap}
\end{figure}

\begin{figure}
	\includegraphics[scale=1.0]{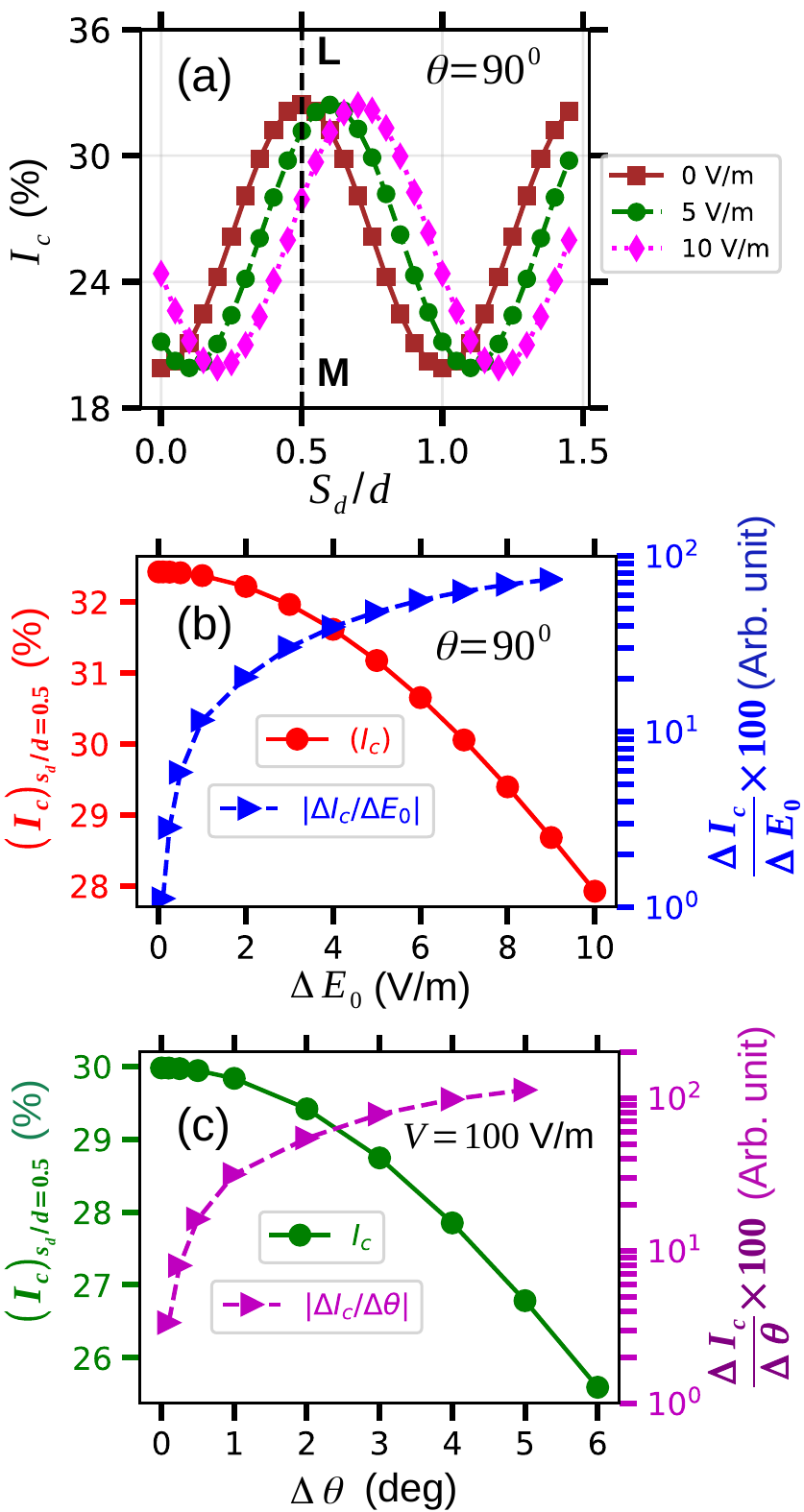}
	\caption{(a) Variation of the intensity ($I_c$) collected by the detector behind the mask grating with the displacement of that grating ($s_d/d$) for different field strengths $E_0$ ($0-10$ V/m) at a fixed angle $\theta=90^o$. The dashed black line (LM) at $s_d/d=0.5$, where $I_c$ for $E_0=0$ V/m attains a peak, suggests that both diffraction and mask grating are in phase. (b) Variation of the intensity $(I_c)_{s_d/d=0.5}$ collected at $s_d/d=0.5$ (left $y$-axis) and its sensitivity to the electric field $(\Delta I_c/\Delta E_0) \times 100$ (right $y$-axis) with $\Delta E_0$. (c) Variation of the intensity $(I_c)_{s_d/d=0.5}$ (left $y$-axis) and its sensitivity to the electric field orientation $(\Delta I_c/\Delta \theta) \times 100$ (right $y$-axis) with $\Delta \theta$.}
	\label{fringe_shift_with_E0_and_theta_plot}
\end{figure}
\subsubsection{Fringe shift}
Being a charged particle, the diffracted wave packet of a proton gets shifted along the transverse direction in the presence of $\vec E$. The variation of the fringe shift ($s$) with $E_0$ and $\theta$ is shown in Fig. \ref{fringe_shift_colormap}. It is observed that the value of $s$ depends upon both $E_0$ and $\theta$. Moreover, Fig. \ref{fringe_shift_colormap} shows that $s$ increases when the electric field is applied in the transverse direction. A maximum shift of $s=2.15\times d$ is obtained for $E_0=100$ V/m and $\theta=90^o$. Furthermore, similar fringe shifts are observed for different combinations of $E_0$ and $\theta$, which is clear from the contour lines as shown in Fig. \ref{fringe_shift_colormap}. This is because the transverse component of the electric field ($E_Y$) that controls the fringe shift, depends upon both $E_0$ and $\theta$, which can be evaluated by $E_Y=E_0 \sin{\theta}$. Next, the sensitivity of the Talbot fringes on $E_0$ and $\theta$ are investigated.

At first, a mask grating with the same opening function ($f=50\%$) as the diffraction grating is placed at the position of the screen ($X_2=2L_T$) as shown in Fig. \ref{diffraction_setup}. After that, the mask grating is translated along the $y$-axis at in steps of $\Delta y=5$ nm, which can be achieved using commercially available electromagnetically actuated nano-positioning stage \cite{WANG2023109753}. The probability of finding the particle behind the mask grating is evaluated by $I_c=\int_{Y=-\infty}^{+\infty}\int_{X=2L_T}^\infty|\psi(X, Y, T)|^2 dX dY$. The variation of the collected intensity ($I_c$) with the displacement of the mask grating ($s_d$) for different field strengths at $\theta=90^o$ is shown in Fig. \ref{fringe_shift_with_E0_and_theta_plot}\textcolor{blue}{(a)}. The figure shows a sinusoidal variation of the intensity with $s_d$, as observed in earlier studies for nanoparticles, such as C\textsubscript{60}, C\textsubscript{60} \cite{PhysRevA.76.013607},  and massive particles, such as fluorous porphyrin having a molecular weight of 10123 amu \cite{eibenberger2013matter}. Initially, at $s_d/d=0$, the mask grating is placed $180^o$ out of phase to the diffraction grating, and a minimum intensity of $I_c=19.9\%$ is observed for $E_0=0$ V/m, as shown in Fig. \ref{fringe_shift_with_E0_and_theta_plot}\textcolor{blue}{(a)}. Thereafter, $I_c$ increases with the displacement of the mask grating for $E_0=0$ V/m, and it becomes maximum ($32.4\%$) at $s_d/d=0.5$ (black dotted line LM) when both the gratings are in phase, and therefore, maximum transmission occurs, as seen in Fig. \ref{fringe_shift_with_E0_and_theta_plot}\textcolor{blue}{(a)}. Furthermore, it can be noticed that the values of the collected intensity $(I_c)_{s_d/d=0.5}$, decreases with the increase in $E_0$ from 0 to 10 V/m due to the fringe shift. Fig. \ref{fringe_shift_with_E0_and_theta_plot}\textcolor{blue}{(b)} shows the variation of $(I_c)_{s_d/d=0.5}$ with the change in electric field strength ($\Delta E_0$), where $(I_c)_{s_d/d=0.5}$ decreases with $\Delta E_0$. To quantify the dependency of $(I_c)_{s_d/d=0.5}$ upon $E_0$, a sensitivity factor is defined as $(SF)_{E_0}=(\Delta I_c/ \Delta E_0)\times 100$, and its variation with $\Delta E_0$ is shown in Fig. \ref{fringe_shift_with_E0_and_theta_plot}\textcolor{blue}{(b)} (right y-axis). It is found that the field sensitivity factor has a value of $(SF)_{E_0}=1.1 \%$, even for small change in the electric field $\Delta E_0=0.1$ V/m, which increases to $73.5\%$ for $\Delta E_0=10$ V/m, as shown in Fig. \ref{fringe_shift_with_E0_and_theta_plot}\textcolor{blue}{(b)}. Therefore, $I_c$ strongly depends on $E_0$, which can be further increased by extending the interaction region beyond $L_T$ (see Fig. \ref{diffraction_setup}), as used in this study. The electric field sensitivity of the Talbot fringe can be used to detect small electric fields.

A similar variation of $I_c$ with $s_d$ is obtained for different orientations ($\theta$) of $\vec E$ (not shown). The variation of $(I_c)_{s_d/d=0.5}$ with $\Delta \theta$ for $E_0=100$ V/m is shown in Fig. \ref{fringe_shift_with_E0_and_theta_plot}\textcolor{blue}{(c)} (left y-axis), where $(I_c)_{s_d/d=0.5}$ decreases with the increase in $\Delta \theta$ due to the increase in fringe shift. Fig. \ref{fringe_shift_with_E0_and_theta_plot}\textcolor{blue}{(c)} shows variation of $\theta$ sensitivity factor $(SF)_{\theta}=(\Delta I_c/ \Delta \theta)\times 100$ with $\Delta \theta$. The figure shows that $(SF)_\theta$ increases from $3.37\%$  to $112.79\%$  while $\Delta \theta$ is increased from $0.1^o$ to $5^o$. Therefore, the Talbot fringe is highly sensitive to the orientation of the applied electric field. Furthermore, $(SF)_\theta$ can be enhanced by extending the electric field interaction region beyond $L_T/2\leq x \leq 3L_T/2$.

\begin{figure}
	\includegraphics[scale=1.0]{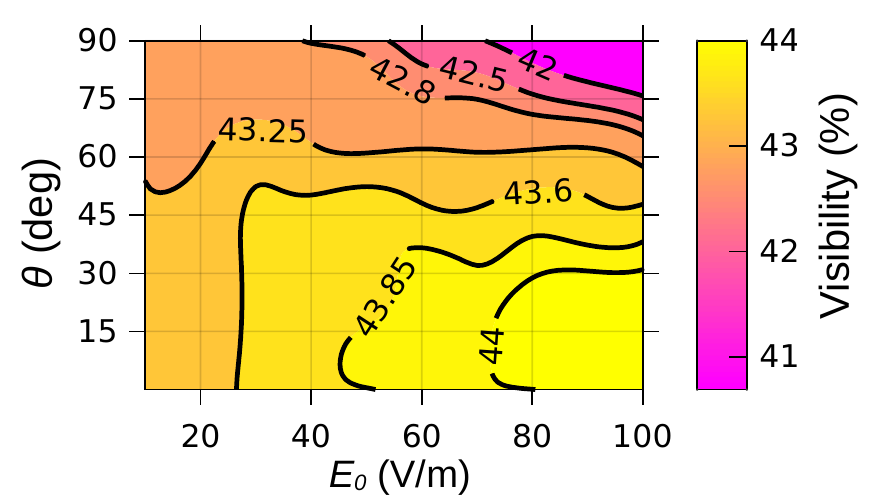}
	\caption{Variation of the fringe visibility ($V$) as a function of magnitude $E_0$ and orientation ($\theta$) of the applied uniform electric field.}
	\label{visibility_plot}
\end{figure}

\subsubsection{Fringe visibility}
The visibility of a diffraction pattern is an important parameter that determines the resolution of the maxima and minima in the fringe. The fringe visibility ($V$) is determined by \cite{Muller_2020}

\begin{equation}
	V=\frac{I_{max} - I_{min}}{I_{max} + I_{min}},
\end{equation}
where $I_{max}$ and $I_{min}$ are the maximal and minimal intensity in the diffraction pattern.

Fig. \ref{visibility_plot} shows the variation of fringe visibility with $E_0$ and $\theta$ at $X_2=2L_T$. It is observed that $V$ depends upon both $E_0$ and $\theta$. Further, it varies from $40.8 \%$ to $44\%$ in the parameter range used in this numerical investigation. A similar value of $V$ ($\sim 40 \%$) has been reported for hydrogen beam diffraction in an earlier experimental study \cite{Muller_2020}. Here, a maximum visibility ($44\%$) is observed at $E_0=100$ V/m and $\theta=0^o$. However, it is seen that the visibility decreases with the increase in $E_0$ when the field is applied in the transverse direction ($\theta=90^o$), as shown in Fig. \ref{visibility_plot}. This is because a strong electric field in the transverse direction destroys the transverse coherence of the wave packet, which is critical to matter wave diffraction. Therefore, the visibility of the fringe pattern decreases with the increase in $E_0$ at $\theta=90^o$. On the other hand, the visibility gets enhanced for strong electric field applied along the direction of propagation ($\theta =0^o$). This happens because, in this case, the electric field accelerates the diffracted wave packet associated with the particle, only in the axial direction ($x$-direction) while the transverse momentum of each peak remains unchanged. Hence, for $\theta=0^o$, the transverse coherence is maintained, and consequently, the visibility increases.

\begin{figure}
	\includegraphics[scale=0.98]{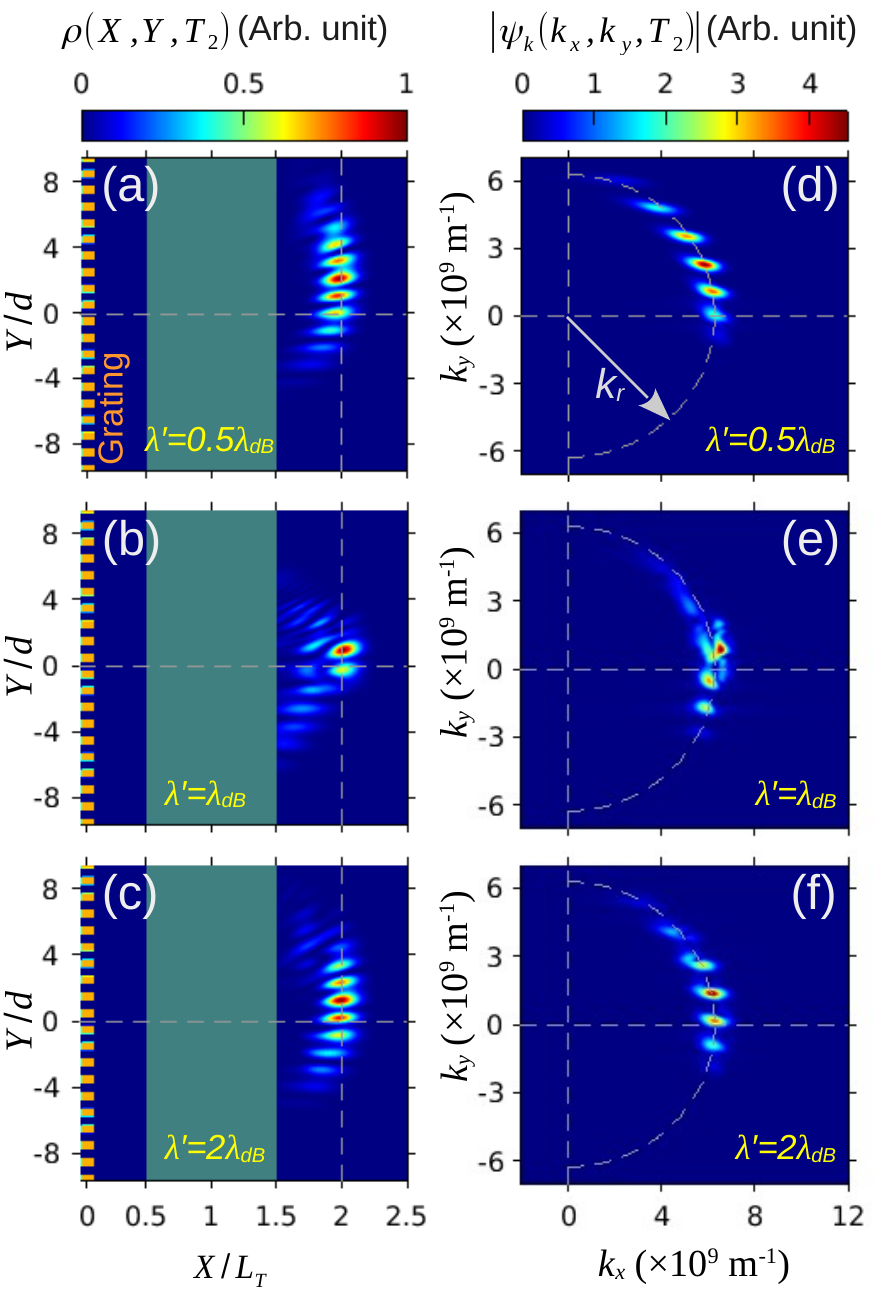}
	\caption{Probability distribution $\rho(X, Y, T)$ of the diffracted wave packet at $T_2=5200$ for spatially modulated electrostatic fields with the length of modulation (a) $\lambda'=0.5 \lambda_{dB}$, (b) $\lambda'=\lambda_{dB}$, and (c) $\lambda'=2 \lambda_{dB}$. The electric field is employed in the transverse direction (along the $y$-axis). The distributions of the corresponding wave packets ($|\psi_k(k_x,k_y,T_2)|$) in the wave number space ($k_x$, $k_y$) are shown for (d) $\lambda'=0.5 \lambda_{dB}$, (e) $\lambda'=\lambda_{dB}$, and (f) $\lambda'=2 \lambda_{dB}$. Here, $k_r$ ($=6.28\times 10^{9}$ m\textsuperscript{-1}) is the wave number of the initial wave packet.}
	\label{plots_for_spatially_modulated_fields}
\end{figure}

\subsection{Near-field diffraction in a spatially modulated electric field}

A spatially modulated electrostatic field is employed in the region $L_T/2\leq x \leq 3L_T/2$ to investigate its effect on the Talbot fringe. The electric field is taken as

\begin{align}
	\vec{E} (x,y)=\hat{y} E_0|\cos(k' x)|,
\end{align}
where $E_0=100$ V/m and the parameter $k' = 2\pi/\lambda'$ determines the degree of spatial modulation of the electric field along the $x$-direction. When the length of modulation $\lambda'$ is comparable to $\lambda_{dB}$, it is expected that the spatial modulation in the electric field will influence the dynamics of the wave packet and hence, on the Talbot pattern, in addition to the fringe shift along the transverse direction that is brought about by $E_0$ alone. To study this effect, in our simulation, the values of $\lambda'$ are taken as $0.5 \lambda_{dB}$ , $\lambda_{dB}$ and 2$\lambda_{dB}$.

Figs. \ref{plots_for_spatially_modulated_fields}\textcolor{blue}{(a)}-\textcolor{blue}{(c)} show the distribution of the probability density $\rho(X,Y,T_2)$ at $T_2=5200$ for $\lambda'=0.5 \lambda_{dB}$, $\lambda_{dB}$, and $2 \lambda_{dB}$, respectively. The distributions of the corresponding wave packets in the wave number space are shown in Figs. \ref{plots_for_spatially_modulated_fields}\textcolor{blue}{(d)}-\textcolor{blue}{(f)}. It can be noticed that for $\lambda'=0.5 \lambda_{dB}$ (see Fig. \ref{plots_for_spatially_modulated_fields}\textcolor{blue}{(a)}), except the fringe shift along the $y$-axis, there seems to be no additional effect on the diffraction pattern, which is similar to the case of a uniform electric field $E_0$ employed along the $y$-axis (cf. Fig. \ref{probability_distribution_xy_kxky_plot}\textcolor{blue}{(c)}). This is also clear from the distribution of $|\psi_k(k_x,k_y, T_2)|$ as shown in Fig. \ref{plots_for_spatially_modulated_fields}\textcolor{blue}{(d)}, which indicates that the modulation in the electric field along $x$-direction has negligible influence on the wave packet when $\lambda'\leq0.5 \lambda_{dB}$.

The wave packet gets affected by the spatial modulation of the electric field when $\lambda'= \lambda_{dB}$, as shown in Fig. \ref{plots_for_spatially_modulated_fields}\textcolor{blue}{(b)}, where distortions in the diffraction pattern become visible. Moreover, the higher order peaks get overlapped each other, which becomes clear from the distribution of $|\psi_k(k_x,k_y, T_2)|$, as shown in Fig. \ref{plots_for_spatially_modulated_fields}\textcolor{blue}{(e)}. This happens because the wave packet, once diffracted by the grating, gets further diffracted by the spatially modulated potential barriers aligned along $y$-direction, which arises due to the spatially modulated electric field. As a result, distortions occur in the final diffraction pattern. As expected, this effect decreases with the increase in $\lambda'$ beyond $\lambda_{dB}$, as shown in Fig. \ref{plots_for_spatially_modulated_fields}\textcolor{blue}{(c)} for $\lambda'=2 \lambda_{dB}$, which is further confirmed from the distribution of $|\psi_k(k_x,k_y,T_2)|$, as shown in Fig. \ref{plots_for_spatially_modulated_fields}\textcolor{blue}{(f)}. Most interestingly, from the semicircular distribution of $|\psi_k(k_x,k_y, T_2)|$ having a radius of $k_r$ ($=6.28 \times 10^{-9}$ m\textsuperscript{-1}), it is clear that the momentum of the diffracted wave packet is conserved in all the cases since the electrostatic field is applied in the transverse direction.

\begin{figure}
	\includegraphics[scale=1.0]{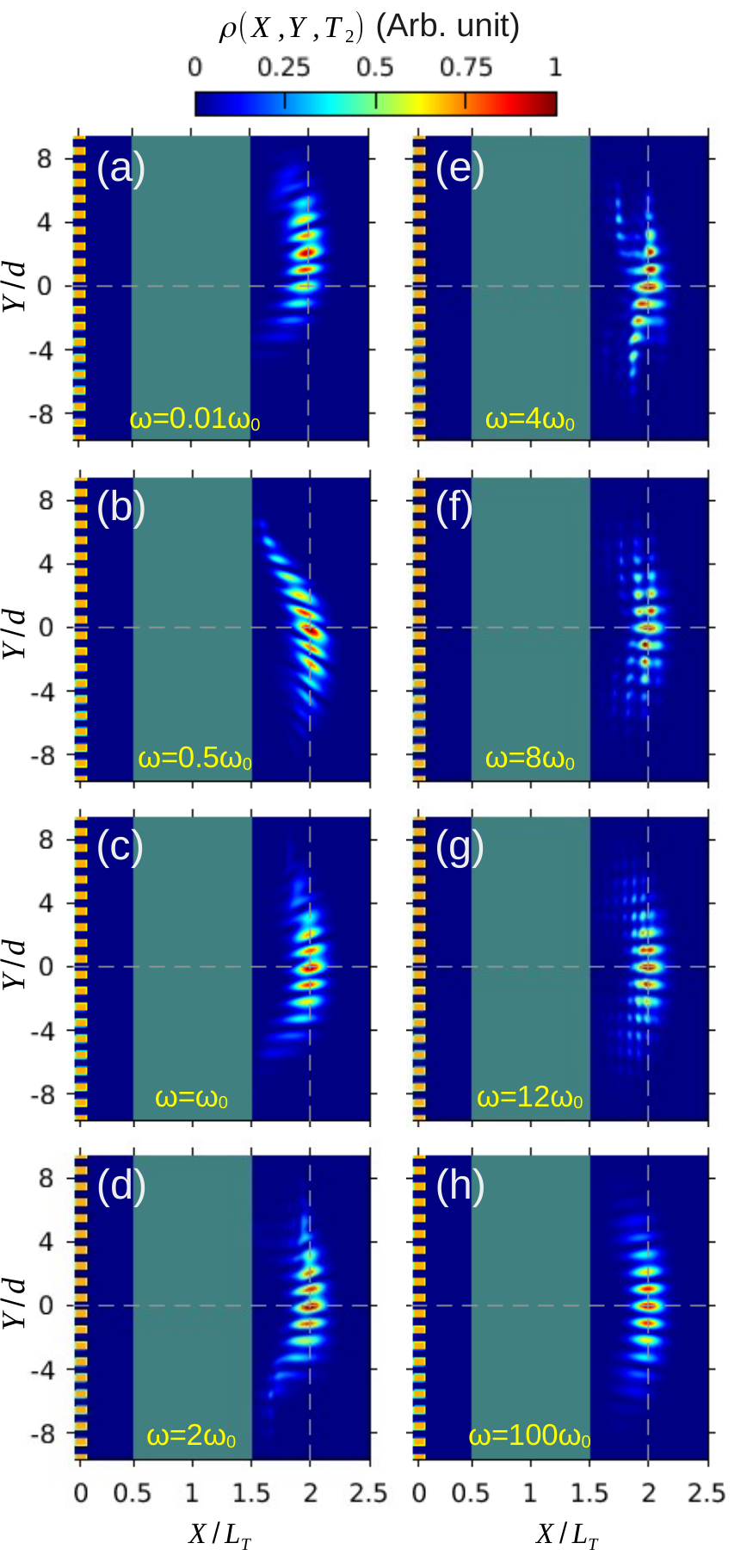}
	\caption{Probability distribution $\rho(X, Y, T_2)=|\psi(X, Y, T_2)|^2$ of the diffracted wave packet after interacting with an oscillating electric field having amplitude $E_0=100$ V/m employed in the region $L_T/2\leq x \leq 3L_T/2$ along the transverse direction ($\theta = 90^0$). The frequency of the oscillating electric fields are (a) $\omega = 0.01 \omega_0$, (b) $\omega = 0.5 \omega_0$, (c) $\omega = \omega_0$, (d) $\omega = 2\omega_0$, (e) $\omega = 4\omega_0$, (f) $\omega = 8\omega_0$, (g) $\omega = 12\omega_0$, and (h) $\omega = 100 \omega_0$, where $\omega_0=2\pi/T_0$ and $T_0$ ($=L_T/\text{v}$) is the time the wave packet takes to cross the electric field region. It can be seen that distortions in $\rho(X, Y, T_2)$ occurs at $\omega=0.5\omega_0$ (panel (b)). With further increase in $\omega$, sidebands started to appear, as shown in panel (d). The sidebands become visible at $\omega=4\omega_0$ (panel (e)). Panel (h) shows that at higher frequencies, such as $\omega=100 \omega_0$, the sidebands disappear.}
	\label{probability_dis_oscillating_fields}
\end{figure}

\begin{figure*}
	\includegraphics[scale=1.0]{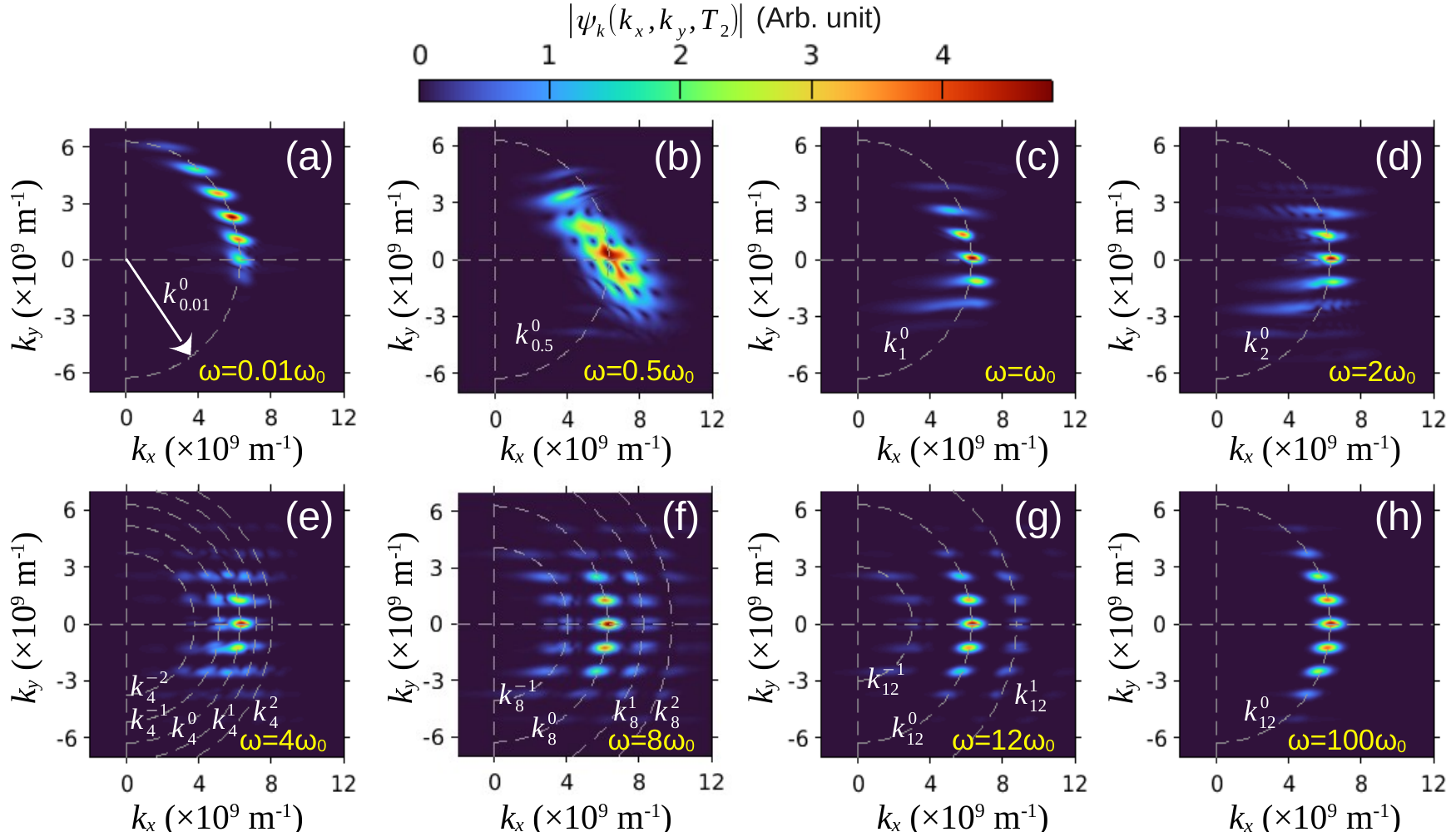}
	\caption{Distribution of the diffracted wave packet $|\psi(k_x,k_y, T_2)|$ in the wave number space ($k_x$, $k_y$) after interacting with the oscillating electric field having a frequency (a) $\omega = 0.01 \omega_0$, (b) $\omega = 0.5 \omega_0$, (c) $\omega = \omega_0$, (d) $\omega = 2\omega_0$, (e) $\omega = 4\omega_0$, (f) $\omega = 8\omega_0$, (g) $\omega = 12\omega_0$, and (h) $\omega = 100 \omega_0$. It can be noticed that when $\omega$ is of the order of $\omega_0$, sidebands start to appear due to the efficient energy transfer from the oscillating field to the wave packet in an integer ($\eta$) multiple of $\hbar \omega$. Positive values of $\eta$ indicate upper band peaks, while negative values of $\eta$ indicate lower band peaks. These sidebands are visible for $\omega=4\omega_0$. The well-formed bands have a circular distribution in the wave number space having radius $k^{\eta}_n$, where $n=\omega/\omega_0$. These sidebands vanish when $\omega<<\omega_0$ or $\omega>>\omega_0$, which is clear from the panels (a) and (h).}
	\label{wavenumber_space_dis_oscillating_fields}
\end{figure*}

\begin{table}[ht]
	\centering
	\caption{Values of the wave numbers ($k^{\eta}_n$) for different sidebands obtained from Eq. \textcolor{blue}({\ref{predicted_k})} and simulation.}
	\label{values_of_wave_numbers}
	\begin{tabular}[t]{lcccc}
		\hline \hline 
		$n$ & Frequency ($\omega$) & Sideband ($\eta$) & \multicolumn{2}{c}{$k^{\eta}_n$ ($\times 10^9$ m\textsuperscript{-1})}\\ \cline{4-5}
		& (rad/s)& & Theory (Eq. \textcolor{blue}({\ref{predicted_k})}) & Simulation\\
		\hline
		$4$ & $4 \omega_0$ & -2 & $3.74$ & 3.71\\
		
		& &$-1$ & $5.16$ & $5.17$\\
		
		& &$0$ & $6.28$ & $6.31$\\
		
		& & $1$ & $7.20$ & $7.23$\\
		
		& & $2$ & $8.02$ & $8.07$\\

		$8$ & $8 \omega_0$ & $-1$ & $3.74 $ & $3.69$\\
		
		& &$0$ & $6.28$ & $6.30$\\
		
		& & $1$ & $8.02$ & $8.01$\\
		
		& & $2$ & $9.46$ & $9.51$\\

		$12$ & $12 \omega_0$ & $-1$ & $2.38$ & $2.41$\\
		
		& &$0$ & $6.28$& $6.30$\\
		
		& & $1$ & $8.52$ & $8.50$\\

		\hline\hline
	\end{tabular}
\end{table}

\subsection{Near-field diffraction in a temporally modulated electric field}

We next investigate the effect of a temporally modulated electric field on the diffraction pattern. The electric field is taken as

\begin{align}
	\vec{E} (x,y,t)=\hat{y} E_0\cos(\omega t),
\end{align}
where $E_0$ ($=100$ V/m) is the amplitude, and $\omega$ is the modulation frequency of the field applied in the transverse direction ($y$-axis) in the region $L_T/2\leq x \leq 3L_T/2$. It is expected that the electric field will influence the dynamics of the diffracted wave packet when the value of $\omega$ is of the order of $\omega_0$ ($=2\pi/T_0 \sim 39.5$ MHz), where $T_0$ ($=L_T/\text{v}$) is the interaction time of the wave packet with the oscillating field and $\text{v}$ is the velocity of the incident particle \cite{steane1995phase}.

Figs. \ref{probability_dis_oscillating_fields}\textcolor{blue}{(a)}-\textcolor{blue}{(h)} show the probability distribution $\rho(X, Y, T_2)$ of the diffracted wave packet after interacting with the temporally modulated electric field for the frequencies ranging from $0.01\omega_0$ to $100\omega_0$. The corresponding distributions of $|\psi_k(k_x,k_y,T_2)|$ in the wave number space ($k_x$, $k_y$) are shown in Figs. \ref{wavenumber_space_dis_oscillating_fields}\textcolor{blue}{(a)}-\textcolor{blue}{(h)}. For $\omega \ll \omega_0$, no additional effects were observed, as shown in Fig. \ref{probability_dis_oscillating_fields}\textcolor{blue}{(a)},  except the fringe shift similar to the case of a uniform electric field (cf. Fig. \ref{probability_distribution_xy_kxky_plot}\textcolor{blue}{(c)}), due to the slower time variation of the field compared to the interaction time $T_0$. Significant distortion occur in $\rho(X,Y,T_2)$ at $\omega=0.5 \omega_0$ (Fig. \ref{probability_dis_oscillating_fields}\textcolor{blue}{(b)}), which is also clear in the distribution of $|\psi_k(k_x,k_y,T_2)|$, as shown in Fig. \ref{wavenumber_space_dis_oscillating_fields}\textcolor{blue}{(b)}. With further increase in $\omega$, sidebands start to appear in the distribution, as shown in Figs. \ref{wavenumber_space_dis_oscillating_fields}\textcolor{blue}{ (c)} and \textcolor{blue}{(d)}. It is observed that, unlike for the uniform field case, the fringe shift in the transverse direction becomes zero when $\omega$ is comparable to $\omega_0$ or more, as shown in Fig. \ref{probability_dis_oscillating_fields}\textcolor{blue}{(c)}-\textcolor{blue}{(h)}. This happens because the wave packet spends an approximately equal amount of time in the oppositely directed electric field of a complete oscillation throughout the interaction. At $\omega=4\omega_0$, sidebands are seem to appear in the diffracted wave packet, as shown in the Fig. \ref{probability_dis_oscillating_fields}\textcolor{blue}{ (e)} and Fig. \ref{wavenumber_space_dis_oscillating_fields}\textcolor{blue}{(e)}. Sideband formation  in matter waves optics has previously been studied for other systems, such as reflection of cold atoms from a vibrating surface \cite{steane1995phase, Herwerth_2013} and scattering of atoms by standing waves of light \cite{bernet2000matter}. However, sideband formation in the Talbot pattern for charged particle matter waves using temporally modulated electric field is observed for the first time in this study. These sidebands appear due to the energy transfer from the oscillating field to the wave packet in an integer multiple of $\hbar \omega$. Therefore, energy of $\eta$\textsuperscript{th} band can be obtained as	$E_\eta=E_{k0} + \eta \hbar \omega$, where $\eta \in \mathbb{Z}$ \cite{Herwerth_2013, byrd2012scattering}. The upper bands are formed when $\eta>0$, and for lower bands, $\eta<0$. Therefore, the resulting momentum of the transmitted wave packet gets discretized for different bands, and the allowed wave numbers can be obtained as
\begin{align}
	k^{\eta}_{n}= \frac{1}{\hbar} \sqrt{2m(E_{k0}+\eta n \hbar \omega_0)},
	\label{predicted_k}
\end{align}
where $n=\omega/\omega_0$. However, it is clear that $\eta$ should be such that $\eta \geq -E_{k0}/n\hbar \omega_0$. The values of the possible wave numbers $k^{\eta}_{n}$ for different bands obtained from Eq. \textcolor{blue}({\ref{predicted_k})} and simulation are listed in Table \ref{values_of_wave_numbers}.

For $\omega=4\omega_0$, several peaks in the upper bands ($k^{1}_{4}$, $k^{2}_{4}$ and $k^{3}_{4}$) and lower bands ($k^{-1}_{4}$, $k^{-2}_{4}$) including the central band ($k^{0}_{4}$) are observed, as shown in Fig. \ref{wavenumber_space_dis_oscillating_fields}\textcolor{blue}{(e)}. From Fig. \ref{wavenumber_space_dis_oscillating_fields}\textcolor{blue}{(e)}, it is clear that the positions of the peaks of different sidebands obtained from the simulation agree well with the predicted values of $k^{\eta}_n$ obtained using Eq. \textcolor{blue}{(\ref{predicted_k})}. Further, an increase in $\omega$ causes the upper and lower bands to move far away from the central one with increasing and decreasing momentum, respectively, as shown in Figs. \ref{wavenumber_space_dis_oscillating_fields}\textcolor{blue}{(f)} and \textcolor{blue}{(g)}. This happens because the energy transfer ($\eta \hbar \omega$) becomes larger at higher frequencies. Moreover, it is observed that at higher frequencies ($\omega> 4\omega_0$), the number of peaks in the upper and lower bands decreases, as shown in Fig. \ref{wavenumber_space_dis_oscillating_fields}\textcolor{blue}{(e)-(g)}. The sidebands completely dissapear at $\omega=100\omega_0$, as shown in Fig. \ref{wavenumber_space_dis_oscillating_fields}\textcolor{blue}{(h)}. This is because the efficiency of energy transfer from the temporally modulated field to the wave packet decreases at higher frequencies. Interestingly, it is observed that the wave number ($k^{\eta}_n$) of the sidebands does not change with the amplitude $E_0$ of the electric field since the energy transfer depends only on $\omega$. Instead, a decrease in $E_0$ causes a reduction in the amplitude of the sidebands. Therefore, the sidebands can be tuned by controlling the modulation frequency ($\omega$) of the employed electric field. 
\section{Conclusion}\label{conclusions}

Influence of (a) uniform, (b) spatially modulated, and (c) temporally modulated electric field ($\vec{E}$) on the near-field diffraction of protons by a nanostructured metallic grating are investigated. To obtain the diffraction pattern, time-domain simulations are performed for two-dimensional Gaussian wave packets by solving the time-dependent Schrödinger's equation using the generalized finite difference time domain (GFDTD-Q) method for quantum systems. The study reveals that:

(i) For a uniform electric field, the Talbot fringes get shifted along the transverse direction, and the amount of fringe shift depends upon $E_0$ and $\theta$. Moreover, the Talbot fringe is sensitive ($(SF)_{E_0}=1.1 \%$ and $(SF)_\theta=3.37 \%$) even to a small $\Delta E_0$ ($=0.1$ V/m) and $\Delta \theta$ ($=0.1^o$). The resolution of $\Delta E_0$ and $\Delta \theta$ can be improved further by extending the region of interaction of the diffracted wave packet with $\vec E$ beyond $L_T$. 

(ii) It is found that the applied uniform electric field can tune the Talbot fringe visibility. 

(iii) For a spatially modulated electric field, significant distortions are observed in the Talbot patterns when the length of modulation ($\lambda'$) is equal to $\lambda_{dB}$. The distortions occur because the wave packet gets further diffracted by the potential barriers arising from the spatial modulation of $\vec{E}$. 

(iv) For a temporally modulated electric field, multiple sidebands are observed for the first time in the Talbot pattern when the frequency of oscillation is of the order of $\omega_0$ ($=2\pi/T_0$), where $T_0$ is the interaction time. The sidebands have discrete momentum states that appear due to an efficient energy transfer from the oscillating electric field to the wave packet at an integer multiple of $\hbar \omega$. Therefore, the momentum states of the sidebands, which are found to be independent of  $E_0$, can be controlled by tuning $\omega$.

The findings of this research will be helpful in various emerging applications of quantum technologies. The electric field sensitivity of the Talbot pattern of protons can be used to develop sensors to accurately measure small electric fields and to align the matter wave optical setup precisely. The electric field-dependent fringe visibility can be used in high-precision and controlled metrology. Finally, sidebands in the Talbot fringe can be used as a precise tool for momentum splitter in matter wave interferometry.

\section*{Acknowledgments}
We gratefully acknowledge the financial support from the funding agency SERB (Department of Science and Technology), Government of India, Grant No. CRG/2022/000112. The simulations were performed at the High Performance Computing facility (HPC2013) at IIT Kanpur.


\nocite{*}
\bibliography{bib_file}

\end{document}